\documentclass{article}
\usepackage{spconfa4,amsmath,graphicx}
\usepackage{hyperref}
\usepackage{cleveref}

\usepackage{cleveref}
\Crefname{equation}{Eq.}{Eqs.}
\Crefname{figure}{Fig.}{Figs.}
\usepackage{caption}
\usepackage{subcaption}

\makeatletter
\def\@fnsymbol#1{\ensuremath{\ifcase#1\or *\or \dagger\or \ddagger\or
   \mathsection\or \mathparagraph\or \|\or **\or \dagger\dagger
   \or \ddagger\ddagger \else\@ctrerr\fi}}
\makeatother

\title{Listen first: Output--Based Multi--Microphone Speech Enhancement}
\name{%
  Panos Apostolidis$^{1,2}$,
  Svend Feldt$^{2}$,
  Zheng-Hua Tan$^{1}$,
  Jan {\O}stergaard$^{1}$,
  Jesper Jensen$^{1,2}$
}
\address{%
  $^{1}$ Aalborg University, Department of Electronic Systems, Aalborg, 9220, Denmark\\
  $^{2}$ Eriksholm Research Centre, Snekkersten, 3070, Denmark\\}

\begin{document}
\ninept
\maketitle
\begin{abstract}
Traditionally, hearing-aid speech enhancement (SE) algorithms rely on \emph{input-based} feature estimation, often derived by a voice activity detector (VAD), to configure beamformers. Yet features extracted from noisy microphone signals can become unreliable in challenging acoustic scenes where users most need help. We introduce a novel paradigm in which the settings of a sound processing system are determined by evaluating characteristics of its \emph{output}. To demonstrate this idea, we employ an \emph{output-based} system that selects among a set of minimum power distortionless response (MPDR) beamformers. Although MPDR beamformers are typically avoided due to their sensitivity to steering errors, we show that they become effective within an output-based framework. We compare the proposed system to a conventional input-based minimum variance distortionless response (MVDR) baseline. Experimental results show that the proposed system consistently outperforms the MVDR baseline, particularly at low SNRs, in terms of SNR, ESTOI and PESQ.
\end{abstract}
\section{Introduction}

\begin{figure}[t]
  \centering
  \begin{subfigure}[b]{0.49\linewidth}
  \centering
  \includegraphics[width=\linewidth]{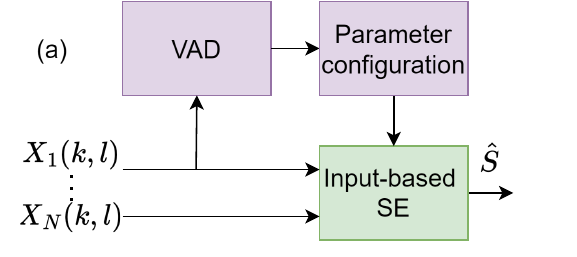}
  \end{subfigure}
  \hfill
  \begin{subfigure}[b]{0.49\linewidth}
  \centering
  \includegraphics[width=\linewidth]{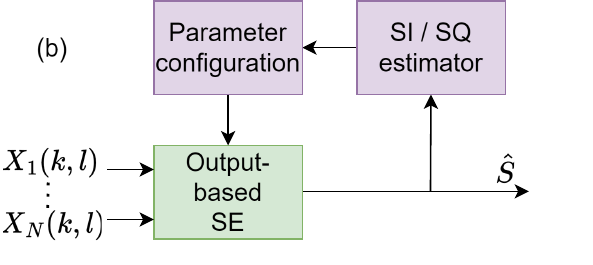}
  \end{subfigure}
  \caption{(a) \emph{input‑based} SE and (b) proposed \emph{output‑based} paradigm.}
  \label{fig:comparison}
\end{figure}

Multi-channel speech enhancement (SE) is widely used in audio applications, including hearing-aid (HA) systems \cite{Poul_beamforming}, which aim to improve speech intelligibility (SI) and sound quality (SQ) by attenuating noise and reverberation \cite{green2022speech}. Conventional HA-oriented SE algorithms, illustrated in \Cref{fig:comparison}a, often rely on acoustic features such as voice-activity-detection (VAD) cues indicating where speech is present in the time-frequency (T-F) domain. Because these features are derived directly from microphone signals, we refer to such systems as \emph{input‑based}. As HAs are low-power devices, the features are often extracted using signal-processing methods \cite{tan2020rvad, gannot2017consolidated}, while deep-learning-based feature extractors have also been explored \cite {heymann2017generic, kim2023dnn}.

In a conventional input-based paradigm, the SE algorithm is decomposed into a multi-channel and a single-channel stage, e.g. a Minimum Variance Distortionless Response (MVDR) beamformer and a post-filter, respectively \cite{jensen2015analysis}. The beamformer relies on estimates of the noise cross-power spectral density and the relative transfer functions (RTFs) from the target to each microphone \cite{gannot2017consolidated}, often provided by a VAD. Since the VAD operates directly on noisy microphone signals, the quality of the estimated speech statistics degrades in challenging acoustic conditions, precisely when HA-users need most support. 

Instead, we propose an alternative paradigm in which a speech-processing system is configured by evaluating the SI or SQ of its \emph{output} rather than extracting features from its input, as illustrated in \Cref{fig:comparison}(b). Related studies have explored output-driven mechanisms \cite{rascon2025direction, kienegger2025self, hafezi2023subspace}. For instance, \cite{rascon2025direction} uses output-based SQ estimates for direction-of-arrival (DOA) correction, while \cite{kienegger2025self} leverages enhanced outputs for target tracking. These approaches modify individual processing components, whereas the proposed paradigm is more general, selecting among any candidate SE system configurations.

As an example of the proposed paradigm, we introduce a SE system that selects from a discrete set of fixed candidate beamformer settings, the one maximizing a speech-intelligibility-related glimpse proportion (GP) measure \cite{cooke2006glimpsing} computed from each candidate's output. This represents an \emph{output‑based} form of processing because all estimates of SI are made after the SE stage, allowing the output itself to guide the settings of the beamforming system. To enable a fair comparison between the proposed system and conventional input-based approaches, we use the same VAD in both systems, ensuring that performance differences arise from the structural distinction between output- and input-based processing rather than from differences in the architecture and complexity of the VAD itself.

In this paper, we demonstrate the potential of the proposed output-based paradigm using a set of candidate minimum power distortionless response (MPDR) beamformers. In particular, we show that the optimal candidate can be reliably selected solely from its output signals, even in low SNR, where input‑based MVDR systems often struggle to estimate speech statistics. Although MPDR beamformers are rarely used due to their sensitivity to steering errors, we find that in an output‑based approach they become effective. Across multiple objective measures, the proposed system consistently outperforms the input‑based baseline. Importantly, the output‑based system retains a performance advantage over the input‑based baseline even under RTF mismatch, highlighting the robustness and practical potential of output‑driven SE for real‑world applications.

\section{Output-based multi-microphone SE system} 

\subsection{Signal model}

We model the noisy microphone signal in the STFT domain as: 

\begin{align}
  \mathbf{X}(k,l) \approx S(k,l)  \mathbf{H}(k) + \mathbf{V}(k,l), 
  \label{equation:eq2}
\end{align}%
where \(S(k,l)\) is the STFT of the target signal, and \(\mathbf{H}(k)\) and \(\mathbf{V}(k,l)\) are the $M$-dimensional  Head-Related Transfer Function (HRTF), and noise vectors \cite{farmani2015maximum}. The indices \(k \in \{0,\dots,K-1\}\) and \(l \in \{0,\dots,L-1\}\) denote frequency bin and time frame, respectively.

\subsection{Output-based MPDR beamforming}

To demonstrate the proposed output-based paradigm, we introduce a beamforming system that selects an MPDR beamformer configuration from a discrete set of candidates based on their output signals. The weights of an MPDR beamformer are given by \cite{capon2005high}

\begin{align}
  \mathbf{W}_{MPDR}(k,l) = \frac{\mathbf{C}_\mathbf{X}^{-1}(k,l) \mathbf{d}_{\theta_i}(k)}{\mathbf{d}_{\theta_i}^H(k)\mathbf{C}_{\mathbf{X}}^{-1}(k,l) \mathbf{d}_{\theta_i}(k)},
\label{equation:eq3}
\end{align}%
where \(\mathbf{C}_\mathbf{X}(k,l) \in C^{M \times M}\) is the covariance matrix of the noisy signal \(\mathbf{X}(k,l)\) and \(\mathbf{d}_{\theta_i}(k)\) denotes the RTF vector with respect to the reference microphone to a target position with direction $\theta_i$.

To enable an output-based MPDR beamformer system, we assume access to a dictionary $\mathbf{d_{\theta}}(k)$ of \(N\) time-invariant candidate RTF vectors \(\mathbf{d}_{\theta_i}(k)\), each corresponding to a candidate target direction \(\theta_i\) at a fixed distance as
\begin{align}
  \mathbf{d_{\theta}}(k) = \{\mathbf{d}_{\theta_1}(k), \mathbf{d}_{\theta_2}(k), ..., \mathbf{d}_{\theta_N}(k)\}.
\end{align}
\label{equation:eq5}%
This dictionary, together with an estimate of \(\mathbf{C}_\mathbf{X}(k,l)\), is used to create candidate MPDR beamformers using \Cref{equation:eq3}, one per candidate target direction, without requiring input-based clean speech or noise statistics. Each candidate MPDR beamformer uses a single direction $\theta_i$ across all frequency bins. The optimal candidate MPDR beamformer is chosen by evaluating the outputs of all candidates and selecting the one that maximizes a performance metric (e.g. a perception-inspired measure, see \Cref{sec:VAD}). This choice of MPDR beamforming is crucial, as it allows each candidate to be constructed without relying on VAD‑based input covariance estimates, enabling purely output‑driven selection among fixed beamformer settings. 

\subsection{Output-based speech intelligibility prediction} 
\label{sec:VAD}

In our output-based system the optimal MPDR beamformer is selected from the candidate set so that a speech-intelligibility-inspired measure is maximized. The evaluation of all candidates is performed on features extracted by a VAD applied to each candidate output.

We express the T-F SNR at the reference microphone as
\begin{align}
  \mathrm{SNR}(k,l)\,
    = 20\log_{10}\left(\frac{|\tilde{S}_{\alpha}(k,l)|}{|V_{\alpha}(k,l)|}\right) \, \, [\mathrm{dB}],
\end{align}
\label{equation:eq6}% 
where \(\tilde{S}_{\alpha}(k,l)\) is the clean target signal at microphone $\alpha$, and \(V_\alpha(k,l)\) is the corresponding noise. A T-F audibility measure $\mathrm{AUD}(k,l)$ is adopted from the Speech Intelligibility Index (SII) \cite{hornsby2004speech,pavlovic2018sii} and defined by clipping the T-F SNR to the range $[-15, 15]$ dB and linearly mapping it to $[0,1]$.

Subsequently, the output-based system includes a SI estimation stage to select the optimal candidate beamformer. For each candidate MPDR beamformer, $\mathrm{AUD}(k,l)$ is estimated from its output signal. For this purpose, we employ a neural VAD, see \Cref{VAD_train}, i.e. a neural network that during inference estimates $\mathrm{AUD}(k,l)$ without access to separated speech or noise signals. Since the VAD produces a per–T‑F audibility estimate $\mathrm{\widehat{AUD}}(k,l)$, we use an intelligibility measure that operates on these audibility patterns. Inspired by the Glimpse Proportion (GP) index \cite{cooke2006glimpsing}, we compute a SI measure as: 

\begin{align}
\mathrm{GP} = \frac{1}{KL} \sum_{k}^{K}\sum_{l}^{L}U({\mathrm{\widehat{AUD}}(k,l) - \mathrm{\gamma}_{\mathrm{GP}})},
\label{equation:eq9}
\end{align}%
where $U(x)$ is the unit step function and $\mathrm{\gamma}_{\mathrm{GP}}$ is a configurable threshold. Essentially, $\mathrm{GP}$ measures the proportion of T-F tiles that contain glimpses of speech, i.e. T-F tiles where the estimated audibility, and thus the SNR, exceeds a selected threshold. Finally, the candidate  whose output yields the highest $\mathrm{GP}$ score is selected as the optimal candidate MPDR beamformer. $\mathrm{GP}$ as defined in \Cref{equation:eq9} is suitable for this purpose because it emphasizes speech‑dominant T–F regions, making it more sensitive to the target direction than measures dominated by noise. In an initial comparison study (omitted here due to space constraints), $\mathrm{GP}$ consistently outperformed other SI and SQ measures estimated from the beamformer outputs.

\begin{figure*}[t]
  \centering
  \includegraphics[width=\textwidth]{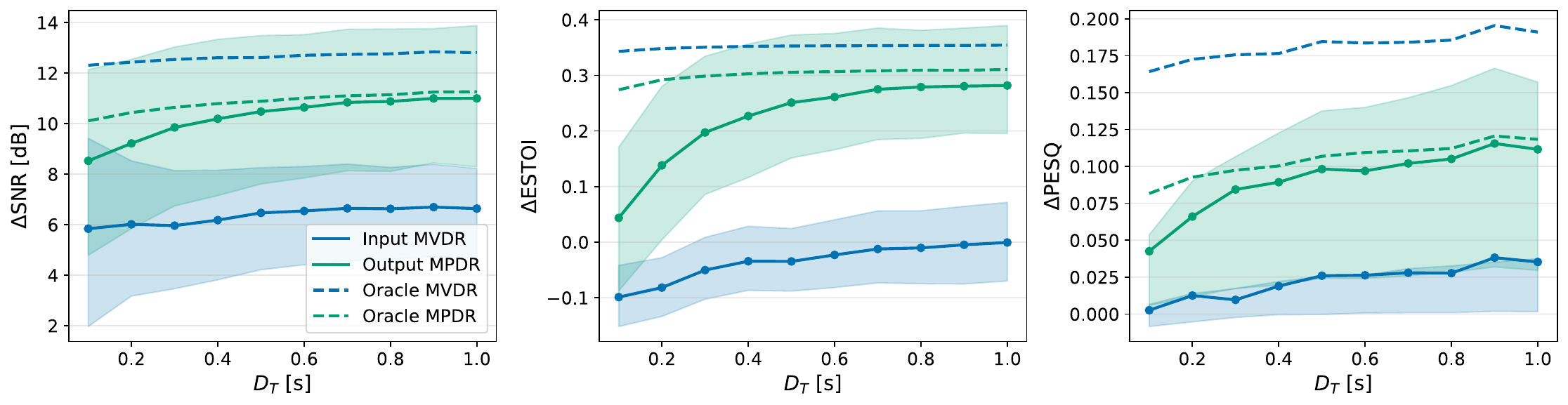}
  {\caption{Performance improvements of input-based and output-based systems with respect to the unprocessed noisy input signal, for $SNR_i = -5 $ dB. Lines correspond to mean performance, while the shading represents 25th and 75th quantiles. }
  \label{fig:matched_full}
  }
\end{figure*}

\begin{figure*}[h]
  \centering
  \includegraphics[width=\textwidth]{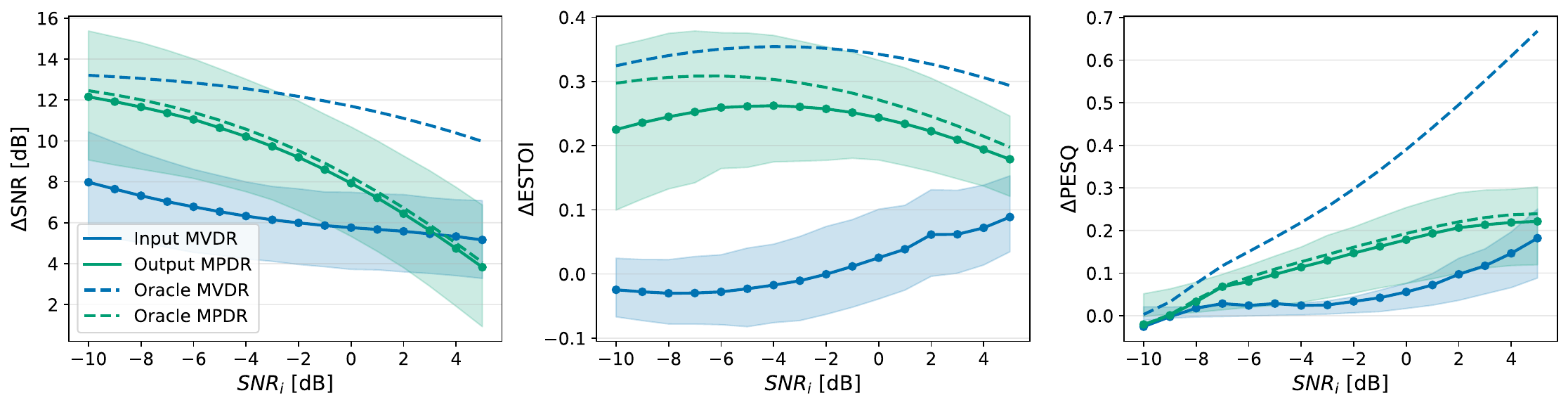}
  {\fontsize{8}{9.6}\selectfont
  \caption{Performance improvements of input-based and output-based systems for $D_T = 0.6$ s.}
  \label{fig:snr_sweep}
  }
\end{figure*}

\section{Input-based MVDR baseline}

As a baseline, we use a conventional input-based MVDR beamformer, which aims to minimize the output noise variance while enforcing a unity gain in the target direction. The weights of the MVDR beamformer are given by \cite{Bitzer2001} 
\begin{align}
  \mathbf{W}_{MVDR}(k,l) = \frac{\mathbf{C}_\mathbf{V}^{-1}(k,l) \mathbf{d}(k)}{\mathbf{d}^H(k)\mathbf{C}_{\mathbf{V}}^{-1}(k,l) \mathbf{d}(k)},
\label{equation:eq10}%
\end{align}
where \(\mathbf{C}_\mathbf{V}(k,l)\) is the noise covariance matrix. 

The MVDR beamformer is equivalent to the MPDR if the estimated RTF \(\mathbf{d}(k)\) is exact and statistics (i.e. \(\mathbf{C}_\mathbf{V}(k,l)\) and \(\mathbf{C}_\mathbf{X}(k,l)\) for MVDR and MPDR beamforming respectively) are known \cite{van2002optimum}. However, if the RTF is mismatched, e.g. due to estimation errors or incorrect target-direction prediction, the MPDR beamformer may cancel the target signal, causing a substantial performance degradation \cite{cox1973resolving}. Thus, practical MPDR performance tends to be worse than MVDR performance, depending on RTF-estimation accuracy.

For this input-based MVDR baseline, we use the same VAD as in \Cref{sec:VAD} to identify speech- and noise-dominated T-F tiles, ensuring a fair comparison. Following a common approach \cite{heymann2017generic, kim2023dnn}, ideal binary masks are formed by applying thresholds to the Audibility function. The ideal binary mask for speech is defined as 
\begin{align}
  \widehat{M}_{S}(k,l) =  \begin{cases}
  0,& \text{if} \quad  \mathrm{\widehat{AUD}}(k,l) \le \mathrm{\gamma}_{S} \\
  1, & \text{if} \quad \mathrm{\widehat{AUD}}(k,l) > \mathrm{\gamma}_{S},
  \end{cases} \label{equation:eq11}
\end{align}
where $\mathrm{\gamma}_S$ is a speech threshold. A noise mask $M_V(k,l)$ is obtained analogously by applying a noise threshold $\gamma_V$.
The speech covariance matrix $\mathbf{C_S}$ can be estimated using, e.g. \cite{heymann2017generic},
\begin{align}
  \mathbf{\widehat{C}_{S}}(k,l) =  \sum_{l=1}^{L}\widehat{M}_{S}(k,l)\mathbf{X}(k,l) \mathbf{X}^H(k,l) .
 \label{equation:eq13-14}
\end{align}%
The noise covariance matrix $\mathbf{\widehat{C}_{V}}(k,l)$ is calculated similarly using the noise mask $\widehat{M}_{V}(k,l)$. Subsequently, the RTF vector is estimated from $\mathbf{\widehat{C}_S}(k,l)$ using the principal eigenvector method \cite{sun2018effect}. The estimated  $\mathbf{{C}_{V}}(k,l)$ and RTF vector $\mathbf{d}(k)$ are then inserted into \Cref{equation:eq10} to compute the MVDR weights.

\section{Experimental setup}

\subsection{Acoustic scene generation}
\label{acoustic_scene}

To generate the source signals used in our simulated acoustic scenes, we use speech signals from the Librispeech Corpus \cite{panayotov2015librispeech}, and point noise sources from a 10-class subset of the ESC-50 dataset \cite{piczak2015esc} of spatially localized sources (e.g. vacuum cleaner).

We then construct acoustic scenes consisting of a HA user, wearing bilateral HAs, i.e. one device on each ear. Each HA is equipped with two microphones, and the left front microphone is arbitrarily selected as the reference microphone. A target talker and one to three point noise sources are placed at random positions on a ring in the horizontal plane of the HAs, with a radius of 1.9 m around the HA user. Isotropic speech-shaped noise (SSN) is also present.

\begin{table*}[h!]
\centering

\begingroup
\fontsize{8}{9.6}\selectfont

\caption{Performance of the input-based MVDR and three output-based MPDR variants under RTF mismatch at $SNR_i=-5$ dB. Bold-faced values indicate significant improvements over the MVDR baseline (Wilcoxon signed-rank test, $p = 0.05$).}
\label{tab:mismatch}

\renewcommand{\arraystretch}{1.15}

\resizebox{\textwidth}{!}{%
\begin{tabular}{
c
|cccc
|cccc
|cccc
}
\hline
\textbf{$D_T$} &
\multicolumn{4}{c|}{\textbf{$\Delta$SNR [dB]}} &
\multicolumn{4}{c|}{\textbf{$\Delta$ESTOI}} &
\multicolumn{4}{c}{\textbf{$\Delta$PESQ}} \\
\hline
 & 
\shortstack{\textbf{Input}\\\textbf{MVDR}} &
\shortstack{\textbf{Output}\\\textbf{MPDR\textsubscript{U}}} &
\shortstack{\textbf{Output}\\\textbf{MPDR\textsubscript{S}}} &
\shortstack{\textbf{Output}\\\textbf{MPDR\textsubscript{F}}} &
\shortstack{\textbf{Input}\\\textbf{MVDR}} &
\shortstack{\textbf{Output}\\\textbf{MPDR\textsubscript{U}}} &
\shortstack{\textbf{Output}\\\textbf{MPDR\textsubscript{S}}} &
\shortstack{\textbf{Output}\\\textbf{MPDR\textsubscript{F}}} &
\shortstack{\textbf{Input}\\\textbf{MVDR}} &
\shortstack{\textbf{Output}\\\textbf{MPDR\textsubscript{U}}} &
\shortstack{\textbf{Output}\\\textbf{MPDR\textsubscript{S}}} &
\shortstack{\textbf{Output}\\\textbf{MPDR\textsubscript{F}}} \\
\hline
0.2 & 6.33 & \textbf{7.40} & \textbf{7.65} & \textbf{9.21} & -0.09 & \textbf{0.02} & \textbf{0.02} & \textbf{0.14} & 0.00 & \textbf{0.02} & \textbf{0.04} & \textbf{0.07} \\
0.4 & 6.42 & \textbf{7.69} & \textbf{7.73} & \textbf{10.19} & -0.04 & \textbf{0.11} & \textbf{0.10} & \textbf{0.23} & 0.02 & \textbf{0.02} & \textbf{0.04} & \textbf{0.09} \\
0.6 & 6.69 & \textbf{7.88} & \textbf{7.87} & \textbf{10.64} & -0.02 & \textbf{0.15} & \textbf{0.14} & \textbf{0.26} & 0.02 & \textbf{0.03} & \textbf{0.03} & \textbf{0.10} \\
0.8 & 6.78 & \textbf{7.95} & \textbf{7.89} & \textbf{10.88} & 0.00 & \textbf{0.17} & \textbf{0.15} & \textbf{0.28} & 0.03 & 0.04 & 0.03 & \textbf{0.11} \\
1.0 & 6.78 & \textbf{7.97} & \textbf{7.90} & \textbf{11.00} & 0.00 & \textbf{0.17} & \textbf{0.16} & \textbf{0.28} & 0.04 & 0.03 & 0.04 & \textbf{0.11} \\
\hline
\end{tabular}%
} % end resizebox

\endgroup
\end{table*}

The HA microphone signals are generated by convolving each clean speech and  point-noise source with the HA Head-Related Impulse Responses (HRIRs) from the OTIMP dataset \cite{moore2019personalized}, which provides HRIRs from 46 individuals with late reverberations removed. For the acoustic scene considered here, we use HRIRs sampled at 48 azimuth angles (7.5° resolution) and at an elevation of 0°.

The convolved signals are summed with isotropic SSN, simulated as a sum of SSN point sources, at the HA microphones. For each mixture, the point‑source noise is level‑adjusted at the reference microphone to be 5–15 dB above the isotropic SSN floor. Each 5‑s utterance uses a new acoustic scene by randomly selecting a HA user (and thus a set of HRIRs), a target location, and one to three point‑noise sources, each with a random position. All signals are processed at a sampling rate of 16 kHz, and the STFT is computed using a 128‑point FFT, an 8 ms Hann window, and a 4 ms hop size.

\subsection{VAD model and training procedure}
\label{VAD_train}

Based on the acoustic scenes described in \Cref{acoustic_scene}, we construct a dataset for training a neural VAD model that estimates $\mathrm{AUD}(k,l)$ for every T-F tile. The total duration of this dataset is 1 hour, which is split into a 90\% train set and 10\% test set. Training segments are 1-second-long and are created by dividing the simulated mixtures into non‑overlapping 1 s excerpts, while the input SNR is uniformly sampled between -15 and 15 dB. For each utterance, a randomly selected microphone $m$ is used, which serves as a form of data augmentation by exposing the VAD to the variability across microphone channels. Subsequently, min-max scaling is applied, and the stacked real and imaginary parts of $X(k,l)$ form the network input.

The neural VAD is implemented as a Convolutional Recurrent Network (CRN) \cite{tan2018convolutional}, combining a convolutional encoder-decoder architecture with an LSTM on the encoder's latent space. The network is trained using the Mean Squared Error (MSE) loss function
\begin{align}
L_{\mathrm{MSE}} = \frac{1}{KL} \sum_{k}^{K}\sum_{l}^{L}(\mathrm{AUD}(k,l) - \mathrm{\widehat{AUD}}(k,l))^2.
\label{equation:eq15}
\end{align}%

The architecture is determined in a hyperparameter tuning stage using Bayesian optimization \cite{dewancker2015bayesian}, yielding a model with 2.9M parameters. Using the resulting architecture, the network is trained for 300 epochs using the Adam optimizer with a learning rate of 0.016, and a batch size of 32. The encoder and decoder contain five causal convolution layers with kernel size $(k,l) = (3,2)$, ELU activations \cite{clevert2015fast}, batch normalization, and a stride of 2 along the frequency axis. Skip connections link the encoder and decoder, while four stacked LSTM layers are inserted between them. A sigmoid activation on the output maps values to $[0,1]$.

\subsection{Implementation details of beamforming systems}
\label{dataset}
 
The beamforming systems are evaluated using the acoustic scenes described in \Cref{acoustic_scene}. The dataset used for beamforming evaluation has a duration of 2 hours and is divided into a validation set and a test set using a 50\%-50\% split. The validation set is used to  tune the hyperparameters $\gamma_S$, $\gamma_V$, and $\gamma_{\mathrm{GP}}$ (defined in \Cref{equation:eq11} and \Cref{equation:eq9}, respectively) with the goal of maximizing the output SNR of both the proposed and the baseline system.

Each utterance is segmented into non-overlapping segments of duration $D_T$, during which beamformer weights remain fixed. As both approaches have access to the full segment before selecting or applying beamformer weights, they are non-causal. A causal implementation would be possible by basing the beamformer selection for each segment on estimates computed from previous segments.

\section{Results}

\subsection{Output-based MPDR vs. input-based MVDR performance} 
\label{matched_performance}

In this section, we evaluate the proposed output-based MPDR system in comparison with the conventional input-based MVDR baseline across different experimental conditions. \Cref{fig:matched_full} shows the performance of the proposed output-based system and input-based baseline as a function of segment duration $D_T$, for an input SNR of $\mathrm{SNR}_i = -5 $ dB. The evaluation is performed in terms of SNR, ESTOI \cite{estoi}, and PESQ \cite{pesq}. Across all metrics and durations, the output‑based MPDR consistently outperforms the input‑based MVDR baseline. A Wilcoxon signed-rank test ($p < 0.05$) was conducted for each performance measure and value of $D_T$ separately, confirming that the observed differences are statistically significant.

The figure also includes the performance of two oracle systems, each representing the upper performance bound for its respective system. The oracle MVDR beamformer has access to the clean speech and noise signals, enabling it to compute the ideal noise covariance matrix $C_V$ and RTF vector. Therefore, this model reflects the maximum performance a time-invariant MVDR beamformer can achieve. In contrast, the oracle MPDR beamformer uses the true RTF corresponding to the target location and thus represents the upper bound for the output-based approach. The performance gap between these two oracle models is mainly due to the duration $D_T$, as for sufficiently long segments the two models' performance converge.

A comparison with the oracle models further illustrates the proposed system's effectiveness. For durations $D_T > 0.5$ s, the performance of the output-based MPDR beamformer approaches that of the oracle MPDR, indicating that the GP-based selection reliably identifies the optimal beamformer even in challenging conditions.

\Cref{fig:snr_sweep} presents the systems' performance for  \(D_T = 0.6 \) s, i.e. relatively slowly-changing beamformer systems, and input SNR values from -10 dB to +5 dB. A Wilcoxon signed-rank test ($p < 0.05$) shows that performance differences are statistically significant for all metrics, except for PESQ, for $\mathrm{SNR}_i \leq -8$ dB. For both SNR and ESTOI, the proposed output-based MPDR beamforming system clearly outperforms the input-based MVDR beamformer, especially at low input SNRs. This behavior reflects the fact that, under such challenging conditions, the input-based VAD struggles to identify speech-dominated T-F tiles correctly, whereas the output-based system can still reliably select the MPDR candidate beamformer pointing towards the correct target direction.

\subsection{Robustness to RTF mismatch}

In \Cref{fig:matched_full} the RTF dictionary was matched, containing that user's HRTFs, including the one matching the target location. To investigate the robustness of the proposed output-based system under realistic conditions, we introduce two mismatches in the RTF dictionary. In the first case, denoted \textbf{MPDR\textsubscript{U}}, the dictionary contains a lower spatial resolution, i.e. RTFs are spaced 15° apart, with the target between entries. In the second case, denoted \textbf{MPDR\textsubscript{S}}, we use non-individualized RTFs measured on a HATS mannequin, meaning that the RTF dictionary does not match the user. For comparison, \textbf{MPDR\textsubscript{F}} denotes the matched condition, where the dictionary includes the true target RTF, corresponding to the results in \Cref{fig:matched_full}.

\Cref{tab:mismatch} shows the performance of the input-based MVDR beamformer and the three output-based MPDR beamformer variants at  $\mathrm{SNR}_i = -5 $ dB. As expected, performance decreases for MPDR\textsubscript{S} and MPDR\textsubscript{U} compared to the fully matched RTFs in MPDR\textsubscript{F}. Nevertheless, despite these mismatches, the output-based MPDR beamformers with mismatched RTFs still outperform the input-based MVDR beamformer in terms of SNR and ESTOI for all values of $D_T$. A Wilcoxon signed-rank test ($p=0.05$) confirms that, for both mismatch conditions, the output‑based MPDR significantly outperforms the input‑based baseline in terms of SNR and ESTOI.

\section{Conclusion}

In this work, we introduced a novel \emph{output‑based} processing paradigm in which a speech‑processing system is configured by evaluating the quality of its \emph{output}, rather than relying on features extracted from its noisy input. We demonstrated this paradigm using a beamforming example, proposing an \emph{output‑based} MPDR system for hearing‑aid applications. Unlike conventional \emph{input‑based} beamforming, which depends on VAD decisions derived from noisy microphone signals, the proposed approach evaluates candidate beamformers directly from their \emph{output}, enabling more reliable decisions in adverse acoustic conditions.
By incorporating a neural VAD trained to estimate an audibility measure, the system can identify the beamformer configuration that maximizes an estimate of speech intelligibility, even at low input SNRs where \emph{input‑based} methods often fail. Moreover, the proposed system maintains its advantage under RTF mismatch conditions, demonstrating robustness when the dictionary is coarse or non‑individualized.

\clearpage

\bibliographystyle{IEEEbib}
\bibliography{refs}

@inproceedings{farmani2015maximum,
  title={Maximum likelihood approach to “informed” sound source localization for hearing aid applications},
  author={Farmani, Mojtaba and Pedersen, Michael Syskind and Tan, Zheng-Hua and Jensen, Jesper},
  booktitle={2015 IEEE international conference on acoustics, speech and signal processing (ICASSP)},
  pages={16--20},
  year={2015},
  organization={IEEE}
}

@Inbook{Bitzer2001,
author="Bitzer, Joerg
and Simmer, K. Uwe",
editor="Brandstein, Michael
and Ward, Darren",
title="Superdirective Microphone Arrays",
bookTitle="Microphone Arrays: Signal Processing Techniques and Applications",
year="2001",
publisher="Springer Berlin Heidelberg",
address="Berlin, Heidelberg",
pages="19--38",
isbn="978-3-662-04619-7",
doi="10.1007/978-3-662-04619-7_2",
url="https://doi.org/10.1007/978-3-662-04619-7_2"
}

@article{capon2005high,
  title={High-resolution frequency-wavenumber spectrum analysis},
  author={Capon, Jack},
  journal={Proceedings of the IEEE},
  volume={57},
  number={8},
  pages={1408--1418},
  year={1969},
  publisher={IEEE}
}

@article{cox1973resolving,
  title={Resolving power and sensitivity to mismatch of optimum array processors},
  author={Cox, Henry},
  journal={The Journal of the acoustical society of America},
  volume={54},
  number={3},
  pages={771--785},
  year={1973},
  publisher={Acoustical Society of America}
}

@article{van2002optimum,
  title={Optimum waveform estimation},
  author={Van Trees, Harry L},
  journal={Optimum Array Processing},
  volume={4},
  pages={428--709},
  year={2002},
  publisher={Wiley}
}

@article{hornsby2004speech,
  title={The Speech Intelligibility Index: What is it and what's it good for?},
  author={Hornsby, Benjamin WY},
  journal={The Hearing Journal},
  volume={57},
  number={10},
  pages={10--17},
  year={2004},
  publisher={LWW}
}

@inproceedings{panayotov2015librispeech,
  title={Librispeech: an asr corpus based on public domain audio books},
  author={Panayotov, Vassil and Chen, Guoguo and Povey, Daniel and Khudanpur, Sanjeev},
  booktitle={2015 IEEE international conference on acoustics, speech and signal processing (ICASSP)},
  pages={5206--5210},
  year={2015},
  organization={IEEE}
}

@article{pavlovic2018sii,
  title={SII—Speech intelligibility index standard: ANSI S3. 5 1997},
  author={Pavlovic, Caslav},
  journal={the Journal of the Acoustical Society of America},
  volume={143},
  number={3\_Supplement},
  pages={1906--1906},
  year={2018},
  publisher={Acoustical Society of America}
}

@article{heymann2017generic,
  title={A generic neural acoustic beamforming architecture for robust multi-channel speech processing},
  author={Heymann, Jahn and Drude, Lukas and Haeb-Umbach, Reinhold},
  journal={Computer Speech \& Language},
  volume={46},
  pages={374--385},
  year={2017},
  publisher={Elsevier}
}

@article{kim2023dnn,
  title={DNN-based Parameter Estimation for MVDR Beamforming and Post-filtering},
  author={Kim, Minseung and Cheong, Sein and Shin, Jong Won},
  journal={Proceedings of the INTERSPEECH, Dublin, Ireland},
  pages={20--24},
  year={2023}
}

@article{cooke2006glimpsing,
  title={A glimpsing model of speech perception in noise},
  author={Cooke, Martin},
  journal={The Journal of the Acoustical Society of America},
  volume={119},
  number={3},
  pages={1562--1573},
  year={2006},
  publisher={AIP Publishing}
}

@article{moore2019personalized,
  title={Personalized signal-independent beamforming for binaural hearing aids},
  author={Moore, Alastair H and de Haan, Jan Mark and Pedersen, Michael Syskind and Naylor, Patrick A and Brookes, Mike and Jensen, Jesper},
  journal={The Journal of the Acoustical Society of America},
  volume={145},
  number={5},
  pages={2971--2981},
  year={2019},
  publisher={AIP Publishing}
}

@inproceedings{piczak2015esc,
  title={ESC: Dataset for environmental sound classification},
  author={Piczak, Karol J},
  booktitle={Proceedings of the 23rd ACM international conference on Multimedia},
  pages={1015--1018},
  year={2015}
}

@inproceedings{tan2018convolutional,
  title={A convolutional recurrent neural network for real-time speech enhancement.},
  author={Tan, Ke and Wang, DeLiang},
  booktitle={Interspeech},
  volume={2018},
  pages={3229--3233},
  year={2018}
}

@article{dewancker2015bayesian,
  title={Bayesian optimization primer},
  author={Dewancker, Ian and McCourt, Michael and Clark, Scott},
  journal={URL https://app. sigopt. com/static/pdf/SigOpt\_ Bayesian\_Optimization\_Primer. pdf},
  year={2015}
}

@article{estoi,
  title={An algorithm for predicting the intelligibility of speech masked by modulated noise maskers},
  author={Jensen, Jesper and Taal, Cees H},
  journal={IEEE/ACM Transactions on Audio, Speech, and Language Processing},
  volume={24},
  number={11},
  pages={2009--2022},
  year={2016},
  publisher={IEEE}
}

@inproceedings{pesq,
  title={Perceptual evaluation of speech quality (PESQ)-a new method for speech quality assessment of telephone networks and codecs},
  author={Rix, Antony W and Beerends, John G and Hollier, Michael P and Hekstra, Andries P},
  booktitle={2001 IEEE international conference on acoustics, speech, and signal processing. Proceedings (Cat. No. 01CH37221)},
  volume={2},
  pages={749--752},
  year={2001},
  organization={IEEE}
}

@article{hafezi2023subspace,
  title={Subspace Hybrid MVDR Beamforming for Augmented Hearing},
  author={Hafezi, Sina and Moore, Alastair H and Guiraud, Pierre H and Naylor, Patrick A and Donley, Jacob and Tourbabin, Vladimir and Lunner, Thomas},
  journal={arXiv preprint arXiv:2311.18689},
  year={2023}
}

@ARTICLE{Poul_beamforming,
  author={Hoang, Poul and de Haan, Jan Mark and Tan, Zheng-Hua and Jensen, Jesper},
  journal={IEEE/ACM Transactions on Audio, Speech, and Language Processing}, 
  title={Multichannel Speech Enhancement With Own Voice-Based Interfering Speech Suppression for Hearing Assistive Devices}, 
  year={2022},
  volume={30},
  number={},
  pages={706-720},
  doi={10.1109/TASLP.2022.3145294}}

@article{green2022speech,
  title={Speech recognition with a hearing-aid processing scheme combining beamforming with mask-informed speech enhancement},
  author={Green, Tim and Hilkhuysen, Gaston and Huckvale, Mark and Rosen, Stuart and Brookes, Mike and Moore, Alastair and Naylor, Patrick and Lightburn, Leo and Xue, Wei},
  journal={Trends in Hearing},
  volume={26},
  pages={23312165211068629},
  year={2022},
  publisher={SAGE Publications Sage CA: Los Angeles, CA}
}

@article{tan2020rvad,
  title={rVAD: An unsupervised segment-based robust voice activity detection method},
  author={Tan, Zheng-Hua and Dehak, Najim and others},
  journal={Computer speech \& language},
  volume={59},
  pages={1--21},
  year={2020},
  publisher={Elsevier}
}

@article{gannot2017consolidated,
  title={A consolidated perspective on multimicrophone speech enhancement and source separation},
  author={Gannot, Sharon and Vincent, Emmanuel and Markovich-Golan, Shmulik and Ozerov, Alexey},
  journal={IEEE/ACM Transactions on Audio, Speech, and Language Processing},
  volume={25},
  number={4},
  pages={692--730},
  year={2017},
  publisher={IEEE}
}

@inproceedings{jensen2015analysis,
  title={Analysis of beamformer directed single-channel noise reduction system for hearing aid applications},
  author={Jensen, Jesper and Pedersen, Michael Syskind},
  booktitle={2015 IEEE International Conference on Acoustics, Speech and Signal Processing (ICASSP)},
  pages={5728--5732},
  year={2015},
  organization={IEEE}
}

@inproceedings{sun2018effect,
  title={Effect of steering vector estimation on MVDR beamformer for noisy speech recognition},
  author={Sun, Xingwei and Wang, Ziteng and Xia, Risheng and Li, Junfeng and Yan, Yonghong},
  booktitle={2018 IEEE 23rd International Conference on Digital Signal Processing (DSP)},
  pages={1--5},
  year={2018},
  organization={IEEE}
}

@article{rascon2025direction,
  title={Direction of arrival correction through speech quality feedback},
  author={Rascon, Caleb},
  journal={Digital Signal Processing},
  volume={158},
  pages={104960},
  year={2025},
  publisher={Elsevier}
}

@article{kienegger2025self,
  title={Self-Steering Deep Non-Linear Spatially Selective Filters for Efficient Extraction of Moving Speakers under Weak Guidance},
  author={Kienegger, Jakob and Mannanova, Alina and Fang, Huajian and Gerkmann, Timo},
  journal={arXiv preprint arXiv:2507.02791},
  year={2025}
}

@article{clevert2015fast,
  title={Fast and accurate deep network learning by exponential linear units (elus)},
  author={Clevert, Djork-Arn{\'e} and Unterthiner, Thomas and Hochreiter, Sepp},
  journal={arXiv preprint arXiv:1511.07289},
  volume={4},
  number={5},
  pages={11},
  year={2015}
}

\end{document}